\documentclass[conference]{IEEEtran}
\IEEEoverridecommandlockouts

\usepackage{cite}
\usepackage{amsmath,amssymb,amsfonts}
\usepackage{graphicx}
\usepackage{algpseudocode}
\usepackage{algorithm}
\usepackage{subfigure}
\usepackage{textcomp}
\usepackage{xcolor}
\def\BibTeX{{\rm B\kern-.05em{\sc i\kern-.025em b}\kern-.08em
    T\kern-.1667em\lower.7ex\hbox{E}\kern-.125emX}}
\usepackage{geometry}
\begin{document}

\title{Integrated Robotic Aerial Base Stations Deployment and Backhaul Design in 6G Multihop Networks\\
}

\author{\IEEEauthorblockN{$\text{Wen Shang}^{\dag}$, $\text{Yuan Liao}^{\dag}$, $\text{Vasilis Friderikos}^{\dag}$, $\text{Halim Yanikomeroglu}^{\S}$}
\IEEEauthorblockA{$^{\dag}$Center of Telecommunication Research, 
King's College London,
London, U.K. \\
$^{\S}$Non-Terrestrial Networks Lab, Department of Systems and Computer Engineering, Carleton University,
Ottawa, Canada\\}
}

\maketitle

\begin{abstract}
To overcome the limited endurance of traditional unmanned aerial vehicles (UAVs), we propose a network of robotic aerial base stations (RABSs) that can energy-efficiently anchor into tall urban landforms, such as lampposts. This approach enables the creation of a hyper-flexible wireless multi-hop network, designed to support green, densified, and dynamic network requirements, thereby ensuring reliable long-term coverage for the whole observed region. The proposed network infrastructure can concurrently address the backhaul link capacity bottleneck and support access link traffic demand in the millimeter-wave (mmWave) frequency band. Specifically, the RABSs grasping locations, resource blocks (RBs) assignment, and route flow control are simultaneously optimized to maximize the served traffic demands. The group of RABSs capitalizes on the fact that traffic distribution varies considerably across both time and space within a given geographical area. Hence, they are able to relocate to suitable locations, i.e., 'follow' the traffic demand as it unfolds to increase the overall network efficiency. To tackle the curse of dimensionality of the proposed mixed-integer problem, we propose a greedy algorithm to obtain a competitive solution with low computational complexity. A wide set of numerical investigations reveals that RABSs could improve the served traffic demand. For instance, compared to networks with randomly deployed fixed small cells, the proposed mode serves at most 65\% more traffic demand.

\end{abstract} 


\section{Introduction}
\vspace{-0.5em}
\IEEEPARstart{W}{ithin} the context of Beyond 5G (B5G) and 6G wireless networks, it is expected that seamless and ubiquitous connectivity should be supported for future dynamic networks with an exponentially increased traffic demand of more than a thousandfold \cite{andreev2019future}.
The aerial base station (ABS), which is mounted on aerial platforms, such as the unmanned aerial vehicle (UAV), has received wide interest recently owing to its flexibility to adapt to dynamic environments and reasonable height to enhance transmission links. 
However, addressing the growing backhaul demands of mobile network topologies with the escalating data traffic is challenging when simply relying on traditional sub-6 GHz wireless backhauling. Moreover, the simultaneous provision of wireless access links to users introduces additional complexity to network management.
  
The promising utilization of millimeter wave (mmWave) bands
\footnote{The band of spectrum with wavelengths between 10 millimeters (30 GHz) and 1 millimeter (300 GHz). Note that this part of the spectrum is also called the extremely high frequency (EHF) band by the International Telecommunication Union (ITU).} 
is expected to address the exponentially increasing traffic demands in future densified heterogeneous networks where a large number of small cells is deployed in conjunction with the macro-BSs \cite{palizban2017automation}.
This potential is rooted in the fact that mmWave bands offer a substantial amount of available spectrum resources, capable of delivering high-throughput access services. Additionally, the use of a highly directional beamforming technique, which is realized through a small-size mmWave antenna, can significantly reduce interference between adjacent receivers
However, the backhaul link capacity becomes another bottleneck for the densified network's service capability. To resolve this challenge, the concept of an integrated access and backhaul (IAB) architecture for 5G cellular networks, has been introduced. Proposed by 3GPP \cite{3gpp.38.174}, this architecture is designed to jointly support wireless backhaul connections and access services with the same infrastructure, further reducing hardware and operation costs \cite{polese2020integrated}. 
To provide fiber-like services for future densified networks, the effective management of shared frequency resources for both backhaul and access links is important.

Recently, due to the flexible nature of ABSs-assisted networks and high deployment expenditure of wired backhaul links, IAB wireless networks have received increasing levels of attention in ABSs-assisted networks \cite{fotouhi2019joint,zhang2023deployment,wang2023deep,zhang2023packet}, from both industry and academia with researchers studying different aspects of network optimization under this proposed network architecture. 
Work in \cite{fotouhi2019joint} optimized UAV's trajectory in an IAB network to maximize the achievable data rate. Overall deployment cost was minimized through the simultaneous optimization of UAV hovering position and backhauling routing by employing deep reinforcement learning algorithms \cite{zhang2023deployment}. In another work \cite{wang2023deep}, to adapt to real-time changes in user demands and environment conditions, the optimal locations of UAVs were determined by employing the dueling double deep Q-network.
However, the spectrum resource allocation between access and backhaul links is predefined and not optimized in studies \cite{fotouhi2019joint, zhang2023deployment,wang2023deep}.
Additionally, the analysis of mean packet-level throughput and energy efficiency of UAV-assisted IAB mmWave networks was investigated in work\cite{zhang2023packet}, with consideration of spectrum allocation between access and backhaul links.  
Numerical results demonstrated that, by adjusting key parameters such as UAV deployment density, and subchannel allocation, network performance could be significantly improved. However, the UAVs' locations were randomly distributed and not well investigated.

Despite many benefits that ABS-assisted wireless networks offer over terrestrial small cells based networks, several challenges remain. One of the main challenges is the constrained service time due to limited onboard battery capacity. A number of efforts have been devoted to overcoming the ABS endurance issue. Firstly, energy-efficient resource management and flying trajectories can be designed separately or jointly to reduce flying energy and thus overcome endurance issues \cite{zhang2020energyefficient}. In addition to that, in the UAV-assisted networks, the propulsion power consumption generated by flying and hovering is much higher than communication consumption, i.e., hundreds of watts compared to several units of power, novel ABS prototypes are developed to further prolong service time. The tethered UAV can provide continuous service by being connected to the ground control platform by a cable/wire that is capable of data and power transfer \cite{kishk2020aerial,zhang2021tethered}. However, the sacrificed mobility limits its broader utilization potential. UAVs powered by laser or optics beam enable a higher level of flexibility to adapt to dynamic networks, and can theoretically provide uninterrupted service when it is close to charging stations \cite{lahmeri2019stochastic}. Nevertheless, this beam transfer can be greatly affected by environmental factors.
In this paper, to overcome the backhaul bottleneck and prolong service time, we propose a multi-hop robotic aerial base stations (RABSs)-assisted IAB mmWave network. The RABS, which is mounted on a UAV and features a flexible grasping mechanism for anchoring to tall street furniture to eliminate energy consumption that is required for hovering/flying, represents a promising approach to support the future green, dynamic, and densified network requirements. A detailed introduction to RABS can be seen in \cite{friderikos2021airborne}.
By concurrently optimizing the RABSs grasping capabilities, resource blocks (RBs) allocation, and flow control of feasible routes, the served traffic demand of the network is maximized. Specifically, the main contributions of this paper are summarized as follows: 
\begin{itemize}
    \item 
Firstly, we introduce a novel aerial base station solution capable of anchoring\cite{friderikos2021airborne}. Anchoring mitigates the endurance limitation of UAV BSs due to the limited onboard battery capacity. Compared to the massive amount for hovering/flying energy consumption (hundreds of watts), RABS allows long-term network connectivity support due to the energy neutral anchoring  \cite{friderikos2021airborne,liao2023optimal}.
Hence, the optimization of flexible RABS attachment to lampposts offers unique ways of enhancing the network's adaptability to dynamic traffic demands. 
\item
Secondly, compared to conventional UAV-enabled network optimizations that consider a constant (or limitless) backhauling capacity, this work supports both backhaul links capacity requirements and access links demands in mmWave links. RBs allocation is optimized to balance the competing demands, with the objective of maximizing served traffic demand. In addition, flow optimization for the proposed wireless multi-hop enabled network is considered. Different from the uniformly distributed traffic demands assumed in previous works, we adopt a lognormal distribution model to estimate the spatial inhomogeneity traffic in heterogeneous networks.
\item 
Lastly, we propose a greedy algorithm with low computational complexity to provide a near-optimal solution. Simulation results demonstrate the improvements in the network performance and flexibility achieved by the proposed network architecture using RABS compared to nominal fixed small cell deployment.
\end{itemize}
\section{System Model and Problem Formulation}
\vspace{-.5em}
\begin{figure}[!t]
\centering
\setlength{\abovecaptionskip}{-0.1cm}
\includegraphics[width=.95\linewidth]{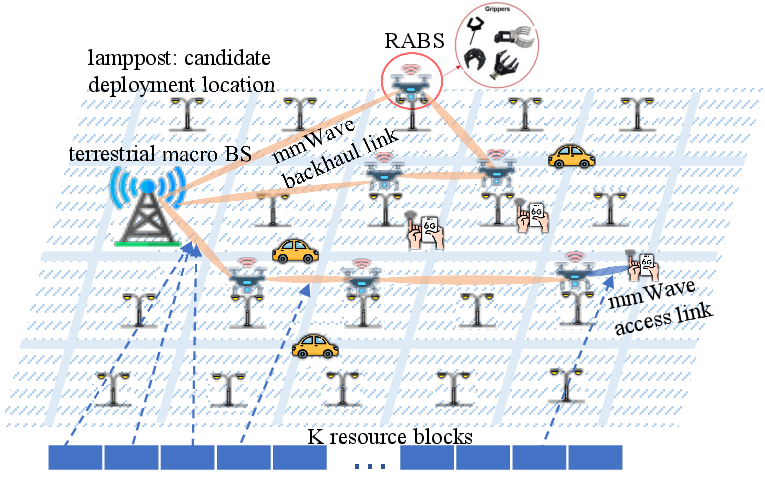}
\caption{Robotic Aerial Base Stations (RABSs) assisted mmWave multi-hop network structure.}
\label{network}
\end{figure}

As shown in Fig. \ref{network}, a downlink RABS-assisted heterogeneous IAB mmWave network is considered in this work, where the RABS is equipped with anchoring end-effectors able to grasp at tall urban landforms and can act as relay nodes. Notably, tall urban landforms such as lampposts provide an almost unlimited resource of potential locations for ‘installing’ small cells and are also deemed highly suitable for 6G small cells given their use of higher frequencies, including mmWave frequency bands. To begin with, a Manhattan-type map that is used for the deployment of the RABS  and associated 6G IAB network in a typical high-density urban environment. This represented as a graph $\mathcal{G} = (\mathcal{V},\mathcal{E})$, where $\mathcal{V}$ is the set of all candidate locations and $\mathcal{E}$ is the set of edges. Each candidate location has a required traffic demand denoted by $D_i, \, \forall i \in \mathcal{V}$. We assume that there are in total $K$ RBs available for both wireless access and backhaul links. The unit wireless backhaul data rate (or capacity) when allocating one RB to the edge $(i,j)$ is denoted by $R^{\text{bh}}_{(i,j)}$. With the same token, the unit wireless access data rate when allocating one RB to the cell $i$ is denoted by $R^{\text{ac}}_{i}$. To provide a feasible solution, the worst-case scenario in terms of achievable data rate is considered, e.g., assuming that the users are distributed at the cell edge.
\subsection{Propagation model}
Both access and backhaul links are expected to operate on the mmWave frequency band. Because the mmWave band is susceptive to obstacles and blockages, Line of Sight (LoS) and Non-Line of Sight (NLoS) channels present different propagation features with distinct parameters. To determine the unit achievable access data rate and backhaul capacity, first, the large-scale path loss for LoS and NLoS links with distance $d$ is considered and expressed as follows\cite{semiari2017inter}, 
\begin{equation}
\begin{aligned}
L(d)&=\mathbb{P}_{\text{LoS}}(d){\text{PL}}_{\text{LoS}}(d)+\big(1-\mathbb{P}_{\text{LoS}}(d)\big){\text{PL}}_{\text{NLoS}}(d)\\
&=\mathbb{P}_{\text{LoS}}(d)\beta d^{-\delta^{\text{LoS}}}+\big(1-\mathbb{P}_{\text{LoS}}(d)\big)\beta d^{-\delta^{\text{NLoS}}},\\
\end{aligned}
\end{equation}
 where $\mathbb{P}_{\text{LoS}}(d)$ is the probability of LoS and it is given by,
\begin{equation}
\mathbb{P}_{\text{LoS}}(d)=\min(18/d,1)\big(1-\exp(-d/36)\big) + \exp(-d/36). 
\end{equation}

${\text{PL}}_{\text{LoS}}(d)$ and ${\text{PL}}_{\text{NLoS}}(d)$ denote the LoS and NLoS path loss with distinct path loss exponents $\delta^{\text{LoS}}$ and $\delta^{\text{NLoS}}$, respectively. Additionally, the term $\beta=(\frac{c}{4\pi f_c})^2$ represents the frequency-dependent free space path loss with 1 m reference distance, and $f_c$ denotes the carrier frequency. 

In this work, the orthogonal frequency division multiplexing (OFDM) is assumed, and each resource block (RB) is constrained to be allocated to at most one transmission link to eliminate interference. For the backhaul links and conditioned to the link state the achievable unit rate capacity when allocating one RB to edge $(i,j)\in\mathcal{E}$ is given by \cite{semiari2017inter},
\begin{equation}
\begin{split}
R_{(i,j)}^{\text{bh}}=w_0\min\Big[{\text{SE}}_{\max},\log_2\big(1+\frac{P_iL(d_{(i,j)})G_{(i,j)}}{w_0N_0}\big)\Big],\\
\end{split}
\end{equation}
where $w_0$ refers to the unit bandwidth of each RB, and ${\text{SE}}_{\max}$ is the maximum spectral efficiency in bps/Hz. $P_i$ denotes the unit transmit power of cell $i$, and a uniformly distributed power over all RBs is assumed, without loss of generality, to reduce the complexity in this work. $w_0N_0$ refers to the noise power, which is the additive white Gaussian noise at the receiver. $L(d_{(i,j)})$ refers to the path loss with backhaul Euclidean distance $d_{(i,j)}$ for edge $(i,j)$. Additionally, to enhance the transmission link and partially compensate for the channel loss, directional beamforming with beam alignment is assumed in this work. For a desired link, the corresponding antenna gain is denoted as $G_{(i,j)}=G^2$, where $G$ refers to the main lobe gain of the base station\cite{semiari2017inter}. 

Similarly to backhaul links propagation approximation, the unit achievable data rate for access links when allocating one RB to cell $i$ is given as follows,
\begin{equation}
R_i^{\text{ac}}=w_0\log_2\big(1+\frac{P_iL(d_i)G_i}{w_0N_0}\big),\;\forall i\in\mathcal{V},
\end{equation}
where $L(d_i)$ represents the path loss of Euclidean distance $d_i$ for wireless access links. Without loss of generality, to provide a robust feasible solution even though users or terminals within this area keep moving, in this work, $d_i$ is calculated when users are distributed at the edge of the cell coverage to obtain a lower bound of the data rate. This scenario represents the worst case in terms of the overall achievable data rate in the network. Additionally, the antenna gain for the access links is denoted as $G_i=G$.

\subsection{Traffic demand distribution model}
\begin{figure}[ht]
\centering
\includegraphics[width=0.8\linewidth]{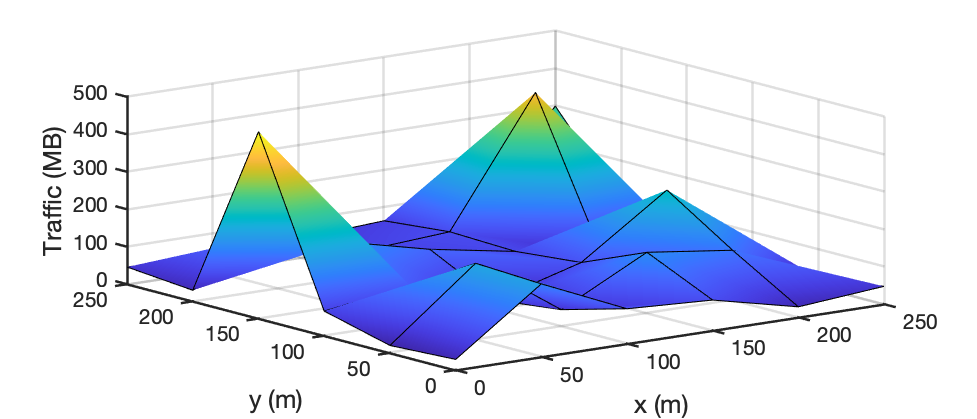}
\caption{Visulaization of spatial traffic distribution.}
\vspace{-0.5em}
\label{spatial_distribution}
\end{figure}

It is well known and widely accepted that cellular traffic demand has significant spatio-temporal variability, and the inherent inhomogeneity of users and traffic demand is one of the key features of the heterogeneous networks \cite{wang2015approach}. In this work, we employ the lognormal distribution model to estimate cells' traffic demands for a specific time duration during which the traffic distribution is assumed to remain unchanged. Hereafter, the estimated traffic volume $D_i$ when placing the RABS in a candidate location $i$ is generated by samples following the lognormal distribution\cite{wang2015approach}, 
\begin{equation}
\label{lognormal}
\begin{aligned}
D_i = \text{lognrnd}\big(\log \mu - \frac{1}{2}\sigma^2,\sigma\big),\;\forall i \in \mathcal{V},
\end{aligned} 
\end{equation}
where $\big(\text{lognrnd}(\log \mu\! - \! \frac{1}{2}\sigma^2, \sigma)\big)$ represents the lognormal distribution with mean $(\log \mu \! - \! \frac{1}{2}\sigma^2)$ and standard deviation $\sigma$, calculated by the natural logarithm. Specifically, component $\mu$ is the mean value and corresponds to a defined time period and use case scenario. In this work, to simplify the complexity, the average value $\mu$ is considered a constant value from \cite{wang2015approach}, and is assumed that it remains unchanged for a specific observation time. Figure \ref{spatial_distribution} shows the inhomogeneous traffic distribution in the spatial domain when setting $\sigma=1$. 
\subsection{Problem formulation and proposed optimization strategy}
Because the hop constraints can significantly increase the overall complexity of the optimization problem, we follow an approach similar to the work in\cite{mcmenamy2019hop}. In this case, we first traverse all potential routes satisfying the hop constraints, flowing from a candidate location and terminating at a fixed macro BS, and denote the set of all potential routes as $\mathcal{P}$. Two subsets of $\mathcal{P}$ are used in the following formulation, $\mathcal{P}_{(i,j)} \subseteq \mathcal{P}$ denotes all potential routes including the edge $(i,j)$,  and $\mathcal{P}_{i} \subseteq \mathcal{P}$ indicates all potential routes sourcing from the candidate location $i$. It is worth noting that, establishing a backhauling hop between two candidate locations situated far away is generally less efficient. This is because the mmWave frequency band is susceptible to blockages and obstacles, and the path loss increases substantially with distance. Therefore, the one-hop backhaul communication range of each RABS is limited by a certain distance to improve the resource efficiency \cite{semiari2017inter}. Hereafter, to set the limited one-hop backhaul distance, the pass loss threshold is constrained to be 150 dB.

To formulate the proposed optimization problem, the binary variable $x_i \in \{0,1\}$ indicates if the candidate location $i$ is selected to deploy a RABS. The integer variables $\{y_{(i,j)}\}$ and $\{z_i\}$ denote the number of RBs allocated to the edge $(i,j)$ and the cell $i$, respectively. Apart from that, the continuous variable $\{f_p\}$ denotes the amount of traffic flow in the route $p$. We then have the following problem formulation,
\begin{subequations}
\label{optimi_prob}
\begin{align}
\; \max_{\mathbf{X},{\mathbf{Y}},{\mathbf{Z},},\mathbf{F} }\; \quad& \sum_{p \in \mathcal{P}} f_p \\
s.t.\quad
\; & \sum_{p \in \mathcal{P}^{(i,j)}} f_p \leq x_i \cdot y_{(i,j)}R^{bh}_{(i,j)}, \; \forall (i,j) \in \mathcal{E}, \label{capacity_con_i} \\
\; & \sum_{p \in \mathcal{P}_{(i,j)}} f_p \leq x_j \cdot y_{(i,j)}R^{bh}_{(i,j)}, \; \forall (i,j) \in \mathcal{E}, \label{capacity_con_j} \\
\; & \sum_{p \in \mathcal{P}_i} f_p \leq x_i \cdot z_i R^{ac}_{i}, \; \forall i \in \mathcal{V}, \label{access_con} \\
\; & z_i R^{ac}_{i} \leq D_i, \; \forall i \in \mathcal{V}, \label{demand_con} \\
\; & \sum_{i \in \mathcal{V}} x_i \leq N, \label{RABS_con} \\
\; & \sum_{(i,j) \in \mathcal{E}} y_{(i,j)} + \sum_{i \in \mathcal{V}} z_i \leq K, \label{Resource_con} \\
\; & x_i \in \{0,1\}, \; \forall i \in \mathcal{V}, \label{Binary_con} \\
\; & y_{(i,j)} \in \mathbb{Z}^+, \; \forall (i,j) \in \mathcal{E}, \label{Integer_cony} \\
\; & z_i \in \mathbb{Z}^+, \; \forall i \in \mathcal{V},\label{Integer_conz} \\
\; & f_p \geq 0, \; \forall p \in \mathcal{P},\label{flow_vari} 
\end{align}
\end{subequations}
where ${\mathbf{X}\overset{\Delta}{=}\{x_i\},{\mathbf{Y}}\overset{\Delta}{=}\{y_{(i,j)}\},{\mathbf{Z}}\overset{\Delta}{=}\{z_i\},\mathbf{F}\overset{\Delta}{=}\{f_p\} }$ are the set of variables. The objective function aims to maximize the served traffic demand of the network. The constraints \eqref{capacity_con_i} and \eqref{capacity_con_j} indicate that all traffic flowing through the edge $(i,j)$ cannot exceed its capacity $y_{(i,j)}R^{bh}_{(i,j)}$, and the flow can pass the edge $(i,j)$ only when there are RABSs deployed in both candidate locations $i$ and $j$, respectively. Eq. \eqref{access_con} refers to the access link rate capacity constraint. The constraint \eqref{demand_con} denotes that all traffic sourcing from the candidate location $i$ cannot exceed its traffic demands $D_i$. The constraint \eqref{RABS_con} indicates that at most $N$ RABSs can be deployed, and \eqref{Resource_con} limits that at most $K$ RBs can be allocated in this network. Constraints (6h) to (6k) denote the type for the decision variables.

To solve the problem in (\ref{optimi_prob}), we first tackle the product of binary variables in (\ref{capacity_con_i}) and (\ref{capacity_con_j}). We substitute the product of $x_i\cdot y_{(i,j)}$ and $x_j\cdot y_{(i,j)}$, by an auxiliary variable $\hat{y}_{(i,j)}$, along with the following additional constraints,
\begin{subequations}
\begin{align}
    & \hat{y}_{(i,j)} \leq x_i\cdot K, \quad \forall (i,j) \in \mathcal{E},\label{xi}\\
    & \hat{y}_{(i,j)} \leq x_j\cdot K, \quad \forall (i,j) \in \mathcal{E}.\label{xj}
\end{align}
\end{subequations}

Then, we linearize the constraint (\ref{access_con}) by substituting the product component $x_i\cdot z_i$ with an auxiliary variable $\hat{z}_i$, along with the following additional constraint,
\begin{equation}
    \hat{z}_i\leq x_i\cdot K, \quad\forall i \in \mathcal{V}.\label{zi}
\end{equation}

After linearizing the product constraints in (6b)-(6c), to solve this mixed-integer linear problem (MILP), we initially identify all routes that satisfy the maximum hop constraint by depth-first search. The proposed MILP can be optimally solved by widely used solvers such as Gurobi\footnote{www.gurobi.com}, which effectively handles integer programming problems utilizing the enhanced branch-and-bound technique. However, the computational complexity of the branch-and-bound approach can grow exponentially with the number of variables in the worst case. Additionally, the number of potential paths for all candidate locations can also increase exponentially with the hop constraint.
With undetermined RABS locations, the number of candidate paths can be enormous. For instance, in a fully connected graph with V nodes, the upper bound on the number of routes to the MBS within three hops has a complexity of $\mathcal{O}(V^3)$. Therefore, we propose a greedy algorithm to find near-optimal solutions with lower computational complexity. 

It is clear that specifying the locations of RABS significantly reduces the number of candidate paths. In the proposed greedy algorithm, the RABS locations are determined first. The objective is to maximize the served traffic demand; hence, deployment locations of RABSs are chosen to serve candidate nodes with higher demand. To deploy the first RABS, the algorithm identifies reachable candidate nodes within one hop distance from the macro base station and sorts them by their traffic demand. The first RABS is deployed at the node with the highest demand among these candidate nodes. Then the algorithm identifies all reachable candidate nodes from the macro base station that satisfy the hop constraints, with determined RABSs acting as relays. The next RABS is then deployed at the location with the highest demand among these identified candidate nodes. This procedure of identifying candidate nodes and deploying RABS according to traffic demand is repeated until no more RABS are available. The obtained deployment solution is denoted as $\mathbf{X}^\star$. After determining RABS locations, related feasible paths can be identified and are denoted as $P'$, which is considerably smaller in scale than the original path set $P$. Subsequently, integer RB allocation variables are relaxed to continuous variables, and the reformulated relaxed linear problem (LP), given as follows, is solved within polynomial time, 
\begin{subequations}
\label{LP_prob}
\begin{align}
\; \max_{\hat{\mathbf{Y}},\hat{\mathbf{Z},},\mathbf{F} }\; &\sum_{p \in \mathcal{P'}} f_p \\
s.t.\quad
\; & \eqref{demand_con}, \eqref{Resource_con}, \eqref{flow_vari}, \eqref{xi}, \eqref{xj}, \eqref{zi},\\
\; & \sum_{p \in \mathcal{P'}^{(i,j)}} f_p \leq \hat{y}_{(i,j)}R^{bh}_{(i,j)}, \quad \forall (i,j) \in \mathcal{E},\\
\; & \sum_{p \in \mathcal{P'}_i} f_p \leq \hat{z}_i R^{ac}_{i}, \quad \forall i \in \mathcal{V}, \\
\; & \hat{y}_{(i,j)} \in \mathbb{R}^+, \hat{z}_i \in \mathbb{R}^+,\quad \forall (i,j) \in \mathcal{E},\forall i \in \mathcal{V},
\end{align}
\end{subequations}
where the obtained optimal solution of \eqref{LP_prob} with specified RABS locations, $\mathbf{X}^\star$, is denoted as $(\hat{\mathbf{Y}}^{\star},\hat{\mathbf{Z}}^{\star},\mathbf{F}^{\star})$. If $(\hat{\mathbf{Y}}^{\star},\hat{\mathbf{Z}}^{\star})$ are integers, a near-optimal solution for the MILP described in \eqref{optimi_prob} is achieved. Otherwise, the obtained solution is rounded down to $(\lfloor \hat{\mathbf{Y}}^{\star}\rfloor,\lfloor\hat{\mathbf{Z}}^{\star}\rfloor)$. With this determined solution for RABS deployment and RB allocations, the LP problem in \eqref{LP_prob} is then solved to obtain the optimal flow result. The Pseudocode of the proposed greedy algorithm is given in Algorithm \ref{algorithm}.

\begin{table}[!t]
\centering
\caption{Summary of Parameters and Notations}
\label{Notation}
\begin{tabular}{p{5.2 cm}|l}
\hline
\textbf{Parameters} & \textbf{Value}\\
\hline
Carrier frequency and unit RB bandwidth $w_0$ & 73 GHz, 2 MHz \\
Total number of RBs $K$ &100-300\\
Path loss exponent $\delta^{\text{LoS}}$ and $\delta^{\text{NLoS}}$ & 2, 3\\
Path loss threshold & 150 dB \\
Maximum spectral efficiency $\textsf{SE}_{\textrm{max}}$ & 4.8 bps/Hz \cite{mcmenamy2019hop}\\
Main lobe antenna gain $G$ & 20 dB\\
Transmission power & 0.1 W \\
Number of RABSs $N$ & 6 \\
Maximum number of hops $H$ & 3 \\
Noise power spectral efficiency & -174 dBm/Hz\\
\hline
\end{tabular}
\end{table}

\begin{algorithm}[t]
\caption{Greedy RABS Multi-Hop Deployment}
\begin{algorithmic}[1]
    \State Initialize graph of all nodes $\mathcal{G}=(\mathcal{V},\mathcal{E})$ with traffic demand $D$, number of RABSs $N$ and maximum hop $H$, path subset $\mathcal{P}'=\emptyset$. 
    \State Find all potential paths that satisfy hop constraint by the depth-first search approach and denote the paths set as $\mathcal{P}$.
    \For   {$r=1$ to $N$}
      \State  Identify paths from set $\mathcal{P}$ that only relay at nodes where a RABS is deployed. Denote reachable end nodes set as $\mathcal{V}'$, and related paths subset as $\mathcal{P}_j',\forall j\in\mathcal{V}'$; 
            \State\textbf{if} {${\mathcal{V}'}$ is not empty} \textbf{then}
      \State Set $x_i=1$, where $i \gets \text{argmax}_{j \in \mathcal{V'}}(D_j)$. Update $\mathcal{P'}=\mathcal{P'}\cup\mathcal{P}_i'$; 
      \State  \textbf{else} break the loop;
    \EndFor
   \State Solve problem in \eqref{LP_prob} with obtained $\mathbf{X}^\star$ and paths subset $\mathcal{P'}$ to obtain solution $(\hat{\mathbf{Y}}^{\star},\hat{\mathbf{Z}}^{\star},\mathbf{F}^{\star})$.
     \State\textbf{if} {$(\hat{\mathbf{Y}}^{\star},\hat{\mathbf{Z}}^{\star})\in \mathbb{Z}^+$}
      \State Obtain a near optimal solution $(\mathbf{X}^\star,\hat{\mathbf{Y}}^{\star},\hat{\mathbf{Z}}^{\star},\mathbf{F}^{\star})$ for \eqref{optimi_prob}.
       \State  \textbf{else}  Solve problem in \eqref{LP_prob} with $(\mathbf{X}^\star,\mathcal{P}')$ and rounded $(\lfloor \hat{\mathbf{Y}}^{\star}\rfloor,\lfloor\hat{\mathbf{Z}}^{\star}\rfloor)$ to obtain near optimal solution for \eqref{optimi_prob}.
\end{algorithmic} \label{algorithm}
\end{algorithm}

\section{Numerical investigations}

Results presented simulate a 250$\times$250 $\rm m^2$ Manhattan-type urban environment as depicted in Fig. \ref{network}. Street lampposts are evenly distributed every 50 meters and all are considered as 25 candidate locations for RABSs grasping, whilst the macro BS is deployed at the location of the original point. Key notation and parameters used in the numerical investigations are summarized in Table \ref{Notation}, and unless otherwise stated, the simulation results are obtained based on values of the parameters listed in Table \ref{Notation}. Apart from that, in this work, we employ the lognormal traffic distribution model to capture the spatial traffic demand as proposed in \cite{wang2015approach}.
 
\begin{figure}[ht]
\centering 
\subfigure[Performance when setting a different number of RABSs.]{\label{com_RABSs} 
\includegraphics[width=.425\textwidth]{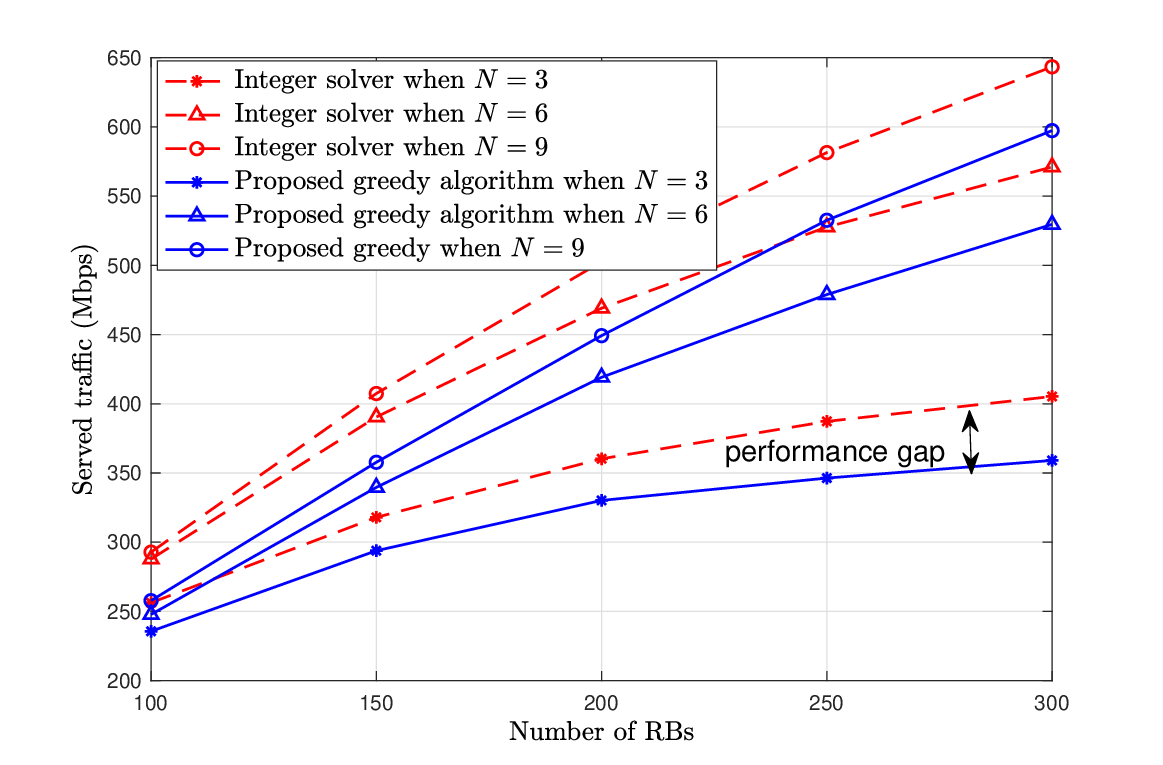}
} 
\subfigure[Performance when setting a different deviation parameter.]{\label{com_sigma}
\includegraphics[width=.425\textwidth]{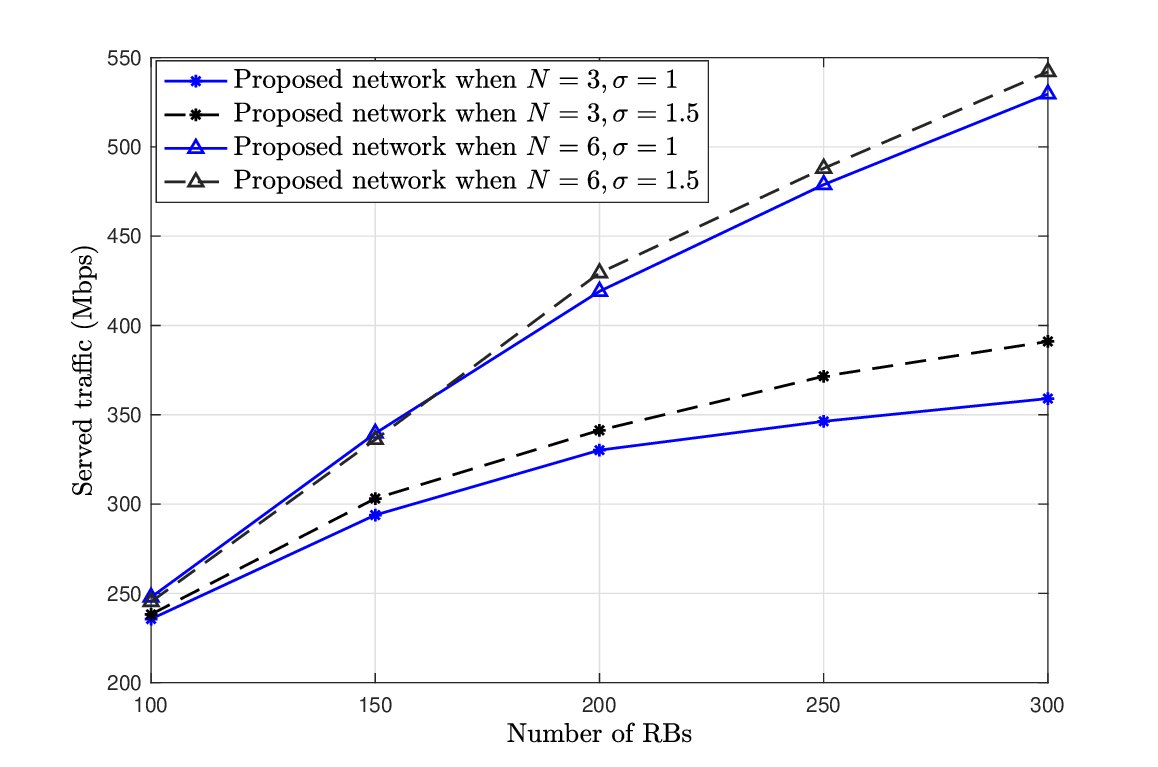}} 
\vspace{-0.5em}
\caption{Performance of the proposed algorithm when setting different $N$, $\sigma$.}
\label{sim_performance} 
\end{figure}

To illustrate the performance of the proposed greedy algorithm, Fig. \ref{sim_performance} compares the optimal solution of served traffic obtained by the integer solver, and the near-optimal solution obtained from the proposed greedy algorithm. First,  observe that the traffic served by RABSs increases monotonically as the number of available RBs increases. However, it has to be noted that, in practice, RABSs may not fully utilize all the RBs due to the power transmission constraint, and this power budget constraint will eventually limit the total served traffic. Second, from Fig. \ref{com_RABSs}, it is evident that the proposed algorithm, with its much lower computational complexity, provides near-optimal solutions with a maximum performance gap of approximately $12\%$ compared to the optimal solutions. Lastly, Fig. \ref{com_sigma} compares the volume of total served traffic under different traffic demand distributions. It is apparent that the served traffic increases with higher deviation values. This difference becomes less pronounced when the number of RABSs is sufficient to cover hotspot traffic demands.

\begin{figure}
\centering 
\subfigure[RB allocation when maximum hop $H = 1$.]{\label{hop1} 
\includegraphics[width=.425\textwidth]{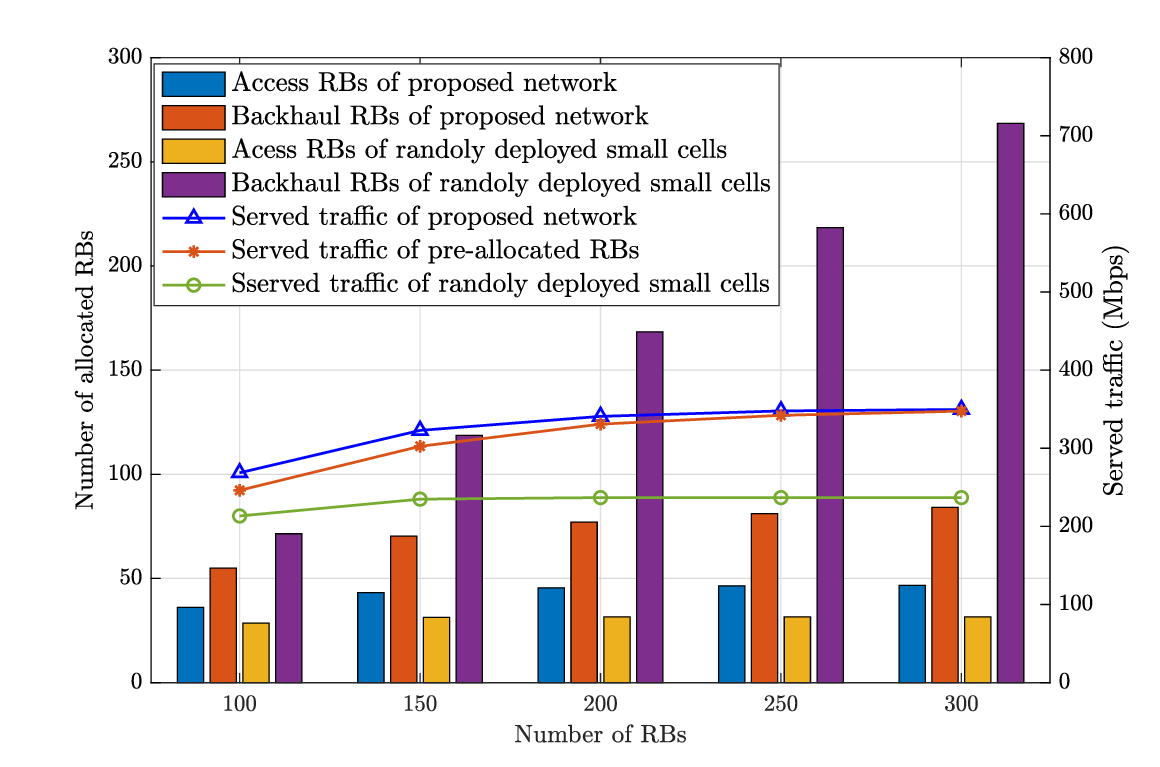}} 
\subfigure[RB allocation when maximum hop $H = 4$.]{\label{hop4}
\includegraphics[width=.425\textwidth]{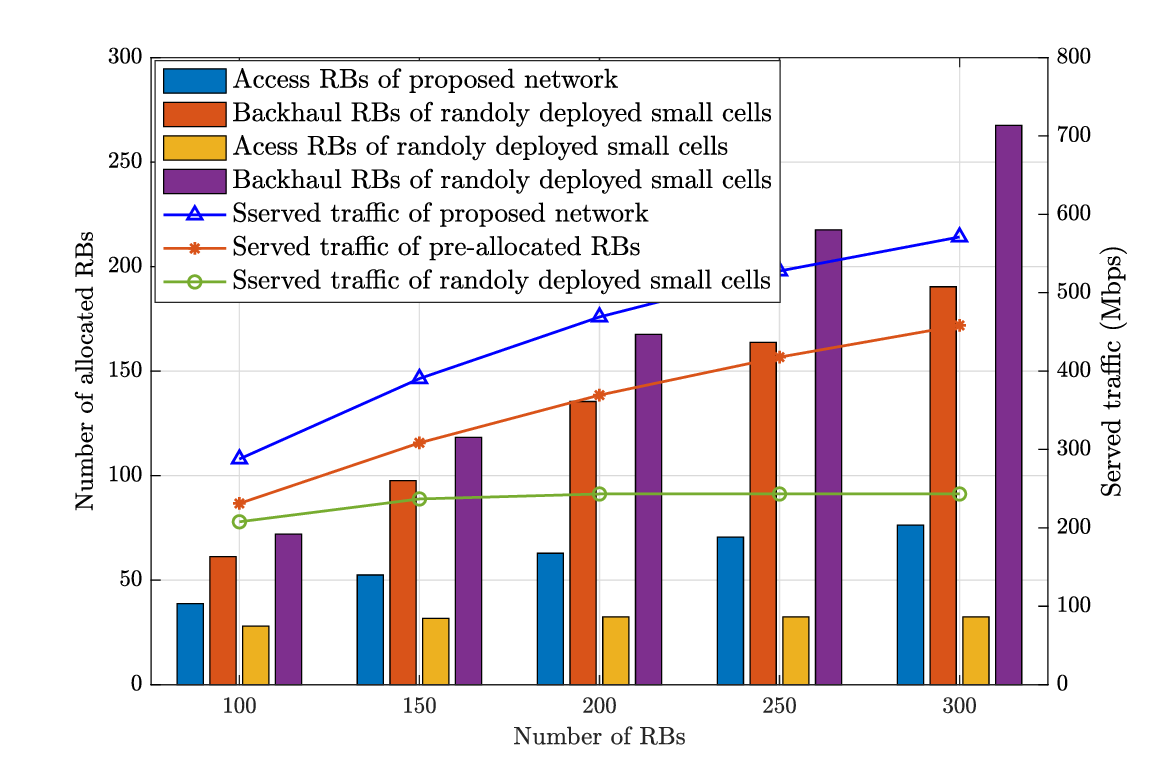}} 
\vspace{-0.5em}
\caption{RB allocation compared with baselines.}
\label{sim_baseline} 
\end{figure}

To illustrate the advantages of the proposed flexible RABS mode, especially when the multi-hop capability is enabled, we compare it with two baseline schemes in Fig. \ref{sim_baseline}. The first baseline involves deploying the same number of fixed small cells, which are assumed to be distributed randomly within the geographical area. The second baseline is the pre-allocated method, where half of the RBs are allocated to wireless access links and the remainder to backhaul links. First, it can be observed that the proposed RABS optimization mode always serves more traffic than the randomly deployed small cells and pre-allocated RBs scenarios, thanks to its controllable and optimal deployment and resource allocation. Second, comparing the volume of total served traffic under different maximum numbers of hops, it can also be observed that the gain from the proposed RABS-assisted network becomes more apparent when the multi-hop capability is enabled. Additionally, when the number of available RBs is sufficiently large, allowing for more hops leads to a more efficient utilization of RBs. Numerically, taking the case of $K = 300$ as an example, the proposed RABSs mode can serve $65\%$ more traffic than fixed small cells when $H = 4$. This gain rate decreases to $30\%$ when $H = 1$, still representing a significant performance improvement. This is because the flexible and optimal deployment of RABSs offers the opportunity to better explore and support hotspots with high traffic demand. This advantage becomes more evident in the network when allowing for more hops. Finally, we compare the proposed flexible IAB RBs allocation strategy with the pre-allocated method. In this use case scenario, it is straightforward to see that the proposed serving mode supports more traffic demand compared to the pre-allocation strategy, and this gain becomes more pronounced as the number of hops increases.

\section{Conclusions}
In this work, to address the pressing need to efficiently support future flexible and high-capacity wireless networks, we propose a RABSs-assisted IAB multi-hop network. The RABSs anchoring location, RBs assignment, and route traffic flow are simultaneously optimized to maximize the network served traffic. A wide set of numerical investigations reveals that the proposed network structure, featuring flexible RABSs capable of anchoring at tall urban landforms depending on the spatial traffic demand, can significantly enhance network performance. For instance, when there are 300 RBs available, compared to networks with randomly deployed fixed small cells, the proposed network with RABSs deployment can serve 65\% more traffic. Additionally, when compared to a static, pre-defined RBs allocation mode, results reveal that the proposed network can improve the network performance through flexible spectrum allocation to strike a balance between access and backhaul links. 

\bibliographystyle{IEEEtran}
 \bibliography{IEEEabrv,reference}

\begin{thebibliography}{10}
\providecommand{\url}[1]{#1}
\csname url@samestyle\endcsname
\providecommand{\newblock}{\relax}
\providecommand{\bibinfo}[2]{#2}
\providecommand{\BIBentrySTDinterwordspacing}{\spaceskip=0pt\relax}
\providecommand{\BIBentryALTinterwordstretchfactor}{4}
\providecommand{\BIBentryALTinterwordspacing}{\spaceskip=\fontdimen2\font plus
\BIBentryALTinterwordstretchfactor\fontdimen3\font minus \fontdimen4\font\relax}
\providecommand{\BIBforeignlanguage}[2]{{%
\expandafter\ifx\csname l@#1\endcsname\relax
\typeout{** WARNING: IEEEtran.bst: No hyphenation pattern has been}%
\typeout{** loaded for the language `#1'. Using the pattern for}%
\typeout{** the default language instead.}%
\else
\language=\csname l@#1\endcsname
\fi
#2}}
\providecommand{\BIBdecl}{\relax}
\BIBdecl

\bibitem{andreev2019future}
S.~Andreev, V.~Petrov, M.~Dohler, and H.~Yanikomeroglu, ``{Future of ultra-dense networks beyond 5G: Harnessing heterogeneous moving cells},'' \emph{{IEEE} Commun. Mag.}, vol.~57, no.~6, pp. 86--92, 2019.

\bibitem{palizban2017automation}
N.~Palizban, S.~Szyszkowicz, and H.~Yanikomeroglu, ``{Automation of millimeter wave network planning for outdoor coverage in dense urban areas using wall-mounted base stations},'' \emph{IEEE Wireless Commun. Lett.}, vol.~6, no.~2, pp. 206--209, 2017.

\bibitem{3gpp.38.174}
3GPP, ``{Integrated Access and Backhaul (IAB) radio transmission and reception},'' {3rd Generation Partnership Project (3GPP)}, Technical Specification (TS) 38.174, 11 2020, version 16.0.0.

\bibitem{polese2020integrated}
M.~Polese \emph{et~al.}, ``Integrated access and backhaul in 5\text{G} mm\text{W}ave networks: Potential and challenges,'' \emph{IEEE Commun. Mag.}, vol.~58, no.~3, pp. 62--68, 2020.

\bibitem{fotouhi2019joint}
A.~Fotouhi \emph{et~al.}, ``Joint optimization of access and backhaul links for \text{UAV}s based on reinforcement learning,'' in \emph{Proc. IEEE Globecom Workshops (GC Wkshps)}, 2019, pp. 1--6.

\bibitem{zhang2023deployment}
Y.~Zhang, M.~A. Kishk, and M.-S. Alouini, ``{Deployment optimization of tethered drone-assisted integrated access and backhaul networks},'' \emph{{IEEE} Trans. Wireless Commun.}, vol.~23, no.~4, pp. 2668--2680, 2024.

\bibitem{wang2023deep}
Y.~Wang and J.~Farooq, ``Deep reinforcement learning based placement for integrated access backhauling in \text{UAV}-assisted wireless networks,'' \emph{IEEE Internet Things J.}, vol.~11, no.~8, pp. 14\,727--14\,738, 2024.

\bibitem{zhang2023packet}
Y.~Zhang \emph{et~al.}, ``{Packet-level throughput analysis and energy efficiency optimization for UAV-assisted IAB heterogeneous cellular networks},'' \emph{{IEEE} Trans. Veh. Technol.}, vol.~72, no.~7, pp. 9511--9526, 2023.

\bibitem{zhang2020energyefficient}
T.~Zhang \emph{et~al.}, ``{Energy-efficient resource allocation and trajectory design for UAV relaying systems},'' \emph{IEEE Trans. Commun.}, vol.~68, no.~10, pp. 6483--6498, 2020.

\bibitem{kishk2020aerial}
M.~Kishk \emph{et~al.}, ``{Aerial base station deployment in 6G cellular networks using tethered drones: The mobility and endurance tradeoff},'' \emph{IEEE Veh. Technol. Mag.}, vol.~15, no.~4, pp. 103--111, 2020.

\bibitem{zhang2021tethered}
S.~Zhang, W.~Liu, and N.~Ansari, ``{On tethered UAV-assisted heterogeneous network},'' \emph{IEEE Trans. Veh. Technol}, vol.~71, no.~1, pp. 975--983, 2021.

\bibitem{lahmeri2019stochastic}
M.-A. Lahmeri, M.~A. Kishk, and M.-S. Alouini, ``{Stochastic geometry-based analysis of airborne base stations with laser-powered UAVs},'' \emph{IEEE Commun. Lett.}, vol.~24, no.~1, pp. 173--177, 2019.

\bibitem{friderikos2021airborne}
V.~Friderikos, ``{Airborne urban microcells with grasping end effectors: A game changer for 6G networks?}'' in \emph{Proc. IEEE Int. Medit. Conf. Commun. Netw. (MeditCom)}, 2021, pp. 336--341.

\bibitem{liao2023optimal}
Y.~Liao and V.~Friderikos, ``{Optimal deployment and operation of robotic aerial 6G small cells with grasping end effectors},'' \emph{{IEEE} Trans. Veh. Technol.}, vol.~72, no.~9, pp. 12\,248--12\,260, 2023.

\bibitem{semiari2017inter}
O.~Semiari, W.~Saad, M.~Bennis, and Z.~Dawy, ``{Inter-operator resource management for millimeter wave multi-hop backhaul networks},'' \emph{{IEEE} Trans. Wireless Commun.}, vol.~16, no.~8, pp. 5258--5272, 2017.

\bibitem{wang2015approach}
{S.\;Wang \textsl{et\;al.}}, ``{An approach for spatial-temporal traffic modeling in mobile cellular networks},'' in \emph{Proc. 27th Int. Teletraffic Congr. (ITC)}, 2015, pp. 203--209.

\bibitem{mcmenamy2019hop}
J.~McMenamy, A.~Narbudowicz, K.~Niotaki, and I.~Macaluso, ``{Hop-constrained mmWave backhaul: maximising the network flow},'' \emph{IEEE Wireless Commun. Lett.}, vol.~9, no.~5, pp. 596--600, 2019.

\end{thebibliography}
\end{document}